# Quantum-Enhanced Diamond Molecular Tension Microscopy for Quantifying Cellular Forces

Feng Xu[1,2,#], Shuxiang Zhang[1,2,#], Linjie Ma[2], Yong Hou[2], Jie Li[3], Andrej Denisenko[4], Zifu Li[5], Joachim Spatz[6,7], Jörg Wrachtrup[4,8], Qiang Wei[1,*], and Zhiqin Chu[2,9,10,*]

1 College of Polymer Science and Engineering, State Key Laboratory of Polymer Materials and Engineering, Sichuan University, Chengdu, 610065, China

2 Department of Electrical and Electronic Engineering, The University of Hong Kong, Hong Kong, China

3 College of Biomass Science and Engineering, Sichuan University, Chengdu, 610065 China

4 3rd Institute of Physics, Research Center SCoPE and IQST, University of Stuttgart, 70569, Stuttgart, Germany

5 National Engineering Research Center for Nanomedicine, College of Life Science and Technology, Huazhong University of Science and Technology, Wuhan, 430074, China

6 Department for Cellular Biophysics, Max Planck Institute for Medical Research, Jahnstraße 29, 69120, Heidelberg, Germany

7 Institute for Molecular Systems Engineering and Advanced Materials (IMSEAM), University of Heidelberg, Im Neuenheimer Feld 225, 69120, Heidelberg, Germany

8 Max Planck Institute for Solid State Research, Stuttgart, Germany

9 School of Biomedical Sciences, The University of Hong Kong, Hong Kong, China

10 Advanced Biomedical Instrumentation Centre, Hong Kong Science Park, Shatin, New Territories, Hong Kong

# F.X. and S.X.Z. contributed equally to this work

* Corresponding to: zqchu@eee.hku.hk (Zhiqin Chu), wei@scu.edu.cn (Qiang Wei)

**Abstract**

The constant interplay and information exchange between cells and their micro-environment are essential to their survival and ability to execute biological functions.

To date, a few leading technologies such as traction force microscopy, have been broadly used in measuring cellular forces. However, the considerable limitations, regarding the sensitivity and ambiguities in data interpretation, are hindering our thorough understanding of mechanobiology. Herein, we propose an innovative approach, namely quantum-enhanced diamond molecular tension microscopy (QDMTM), to precisely quantify the integrin-based cell adhesive forces. Specifically, we construct a force sensing platform by conjugating the magnetic nanotags labeled, force-responsive polymer to the surface of diamond membrane containing nitrogen vacancy (NV) centers. Thus, the coupled mechanical information can be quantified through optical readout of spin relaxation of NV centers modulated by those magnetic nanotags. To validate QDMTM, we have carefully performed corresponding measurements both in control and real cell samples. Particularly, we have obtained the quantitative cellular adhesion force mapping by correlating the measurement with established theoretical model. We anticipate that our method can be routinely used in studying important issues like cell-cell or cell-material interactions and mechanotransduction.

**Introduction**

Biochemical factors in the environment are known to affect living organisms and have been investigated for quite a long time (*1, 2*). Interestingly, recent evidence has also shown that physical cues such as mechanical forces can constantly be generated inside biological systems and get transmitted to their surroundings (*3-5*). The involved mechanical information not only result in deformation and motion but also stimulate physiological functions of lives (*6-8*). Normally, the forces associated with a single cell can range from piconewtons to several nanonewtons, corresponding to molecular and cellular level, respectively (*9, 10*). In this regard, inventing a reliable tool to quantify the mechanical interactions between the cell and substrate, especially at the single cellular level, is crucial for our basic understanding of many important biological processes such as morphogenesis, tissue repair and tumor metastasis (*5, 11*).

Various methods have been successfully developed for measuring cellular adhesive forces in the past few decades. In general, these approaches can be divided into three categories: 1) the first type relies on monitoring the deformation of the substrate to estimate the force, with prime examples being the so-called cellular traction force microscopy (TFM) and micropillar-based force measuring apparatus (*12*); 2) the second category is the single cell force spectroscopy by using an instrument like atomic force microscopy (AFM) or magnetic/optical tweezer systems (*13*); and 3) the third kind is the molecular tension-based fluorescence microscopy (MTFM) or similar tension gauge tether (TGT) systems with the help of force-sensitive fluorophores (*14, 15*). Although these techniques have been well established as standard tools in mechanobiology study, several issues have also been raised during their implementation in actual cellular measurements. For example, the intrinsic experimental caveats of conventional TFM are known to be computationally intensive and, furthermore, such method can mainly sense the shear tractions at nanonewton (nN) level (*16, 17*). In addition, the MTFM and TGT suffers from photo-bleaching of fluorophores with a stochastic nature (*18, 19*). Therefore, the development of a new technique to accurately measure the cell adhesive forces, preferred in a fluorescent label-free manner, is vital to the development of mechanobiology.

The nitrogen vacancy (NV) centers, a kind of photoluminescent defect in diamond, displays a number of attractive features, including unlimited photostability, unique spin properties with optical readout, chemical inertness, flexible modalities, and excellent biocompatibility (*20-22*). Specifically, the electronic-spin-dependent photoluminescence associated with negatively charged NV center (NV$^-$) can facilitate optical readout of various physical quantities (e.g., magnetic fields (*23*), electric fields (*24*), temperature (*25*), etc.) at ambient conditions via conventional fluorescence microscopy, namely quantum sensing. The NV-based nanoscale quantum sensors, hosted in biocompatible and robust diamond materials, show great promise for applications ranging from fundamental to applied sciences (*26-28*). In particular, a set of promising biosensors have been developed in recent years, including intracellular thermometers (*29-33*), intracellular orientation tracking agents (*34, 35*), intracellular free radicals detectors (*36-40*), monitoring of physiological species (*41-43*), detection of neuronal action potential (*44, 45*), magnetic imaging of biomolecules (*46-54*), etc. Despite all these exciting progresses, to the best of our knowledge, the direct sensing of weak mechanical signals in living systems has never been achieved yet.

Here, by combining next-generation quantum measurement platforms, with innovative biointerface engineering technologies, we have developed a novel method, termed quantum-enhanced diamond molecular tension microscopy (QDMTM) for the accurate measurement of cellular adhesion forces (Scheme 1). By coupling mechanical signals to the fluorescence of NV centers, the minute cellular forces induced changes of the polymer will be properly quantified through well-established quantum sensing protocols. We experimentally verified the QDMTM in a series of carefully designed control experiments, and further showcased the semi-quantitative/quantitative mapping of cellular forces at the single cellular level. Our experimental results agreed well with theoretical calculations, suggesting the proposed platform could be in principle upgraded into a standardized toolkit.

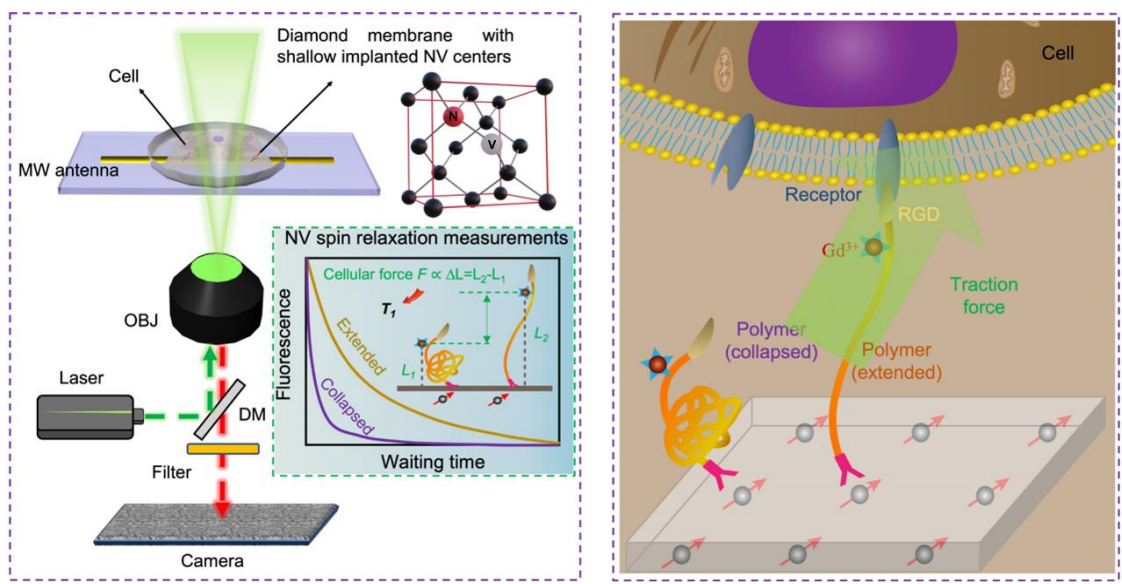

**Scheme 1. Schematic diagram illustrating the design of QDMTM.** The left panel shows the working principle of the widefield quantum diamond microscope. Insert indicates the energy level

of the NV centers. The right panel shows the exact force sensing mechanism. MW antenna: microwave antenna; OBJ: objective; DM: dichroic mirror.

**Results**

**Construction of a robust quantum diamond biosensing platform**

The key of our novel sensing platform, namely the QDMTM, was to correlate (as shown in Scheme 1) NV spin relaxometry with force-induced polymer stretching, which was documented using the well-known Worm-Like Chain model (*55*). In the case where the magnetic labels ($Gd^{3+}$ ions) were attached to the diamond surface through spring-like polymer, the relationship between the relaxation rate $\varGamma_1$ and NV-$Gd^{3+}$ distance $h$ has been known to obey the following relationship (*56*):

$$\varGamma_1 \propto h^{-3} \quad (1)$$

We adopted our previously developed single-crystalline ultrathin diamond membrane (~ 30 µm) with shallow implanted NV centers to work as a widefield quantum sensing substrate. To enable the NV-based measurement of the integrin-based cell adhesive force, such mechanical signals can be converted to magnetic ones using a transducer, i.e., the tailor-made force-responsive polymer (Fig. 1A, Scheme S1 and fig. S1). Specifically, it contained an integrin ligand Cyclo(RGDfK) and a molecular magnet $Gd^{3+}$ on one side, the main molecular chain polyethylene glycol (PEG) serving as the spring element in the middle and a silane anchor on the other end for immobilization (fig. S2). The PEG chain serving an entropic spring was capable of sensing forces on the order of a few tens of pN (*19*) and provided a bioinert background to avoid non-specific interactions with cells (*57*). The deformation of the PEG chain altered the distance between the $Gd^{3+}$ ions and the NV centers, leading to the change of NV spin relaxation time $T_1$ ($1/\varGamma_1$) which can be quantitatively measured (*43*).

Anchoring these force-responsive polymers within the effective sensing range of NV centers (~25 nm above the diamond surface (*58*)), meanwhile, minimizing the thickness of any functionalization layer while retaining excellent surface morphology and coverage is crucial for the $T_1$ test. To enable the conjugation of the designed polymers to the chemically inert surface of diamond, we firstly managed to deposit a stable layer of hybrid silica, and such an interface with high reactivity without further activation was sufficient to enhance intra-layer interactions (*59*) and stabilize the PEG coating. Benefited from this silica layer, the polymer could be easily immobilized onto the diamond surface to construct the desired force sensing platform in mild conditions to avoid detrimental influence on the spin property of NV centers (fig. S3).

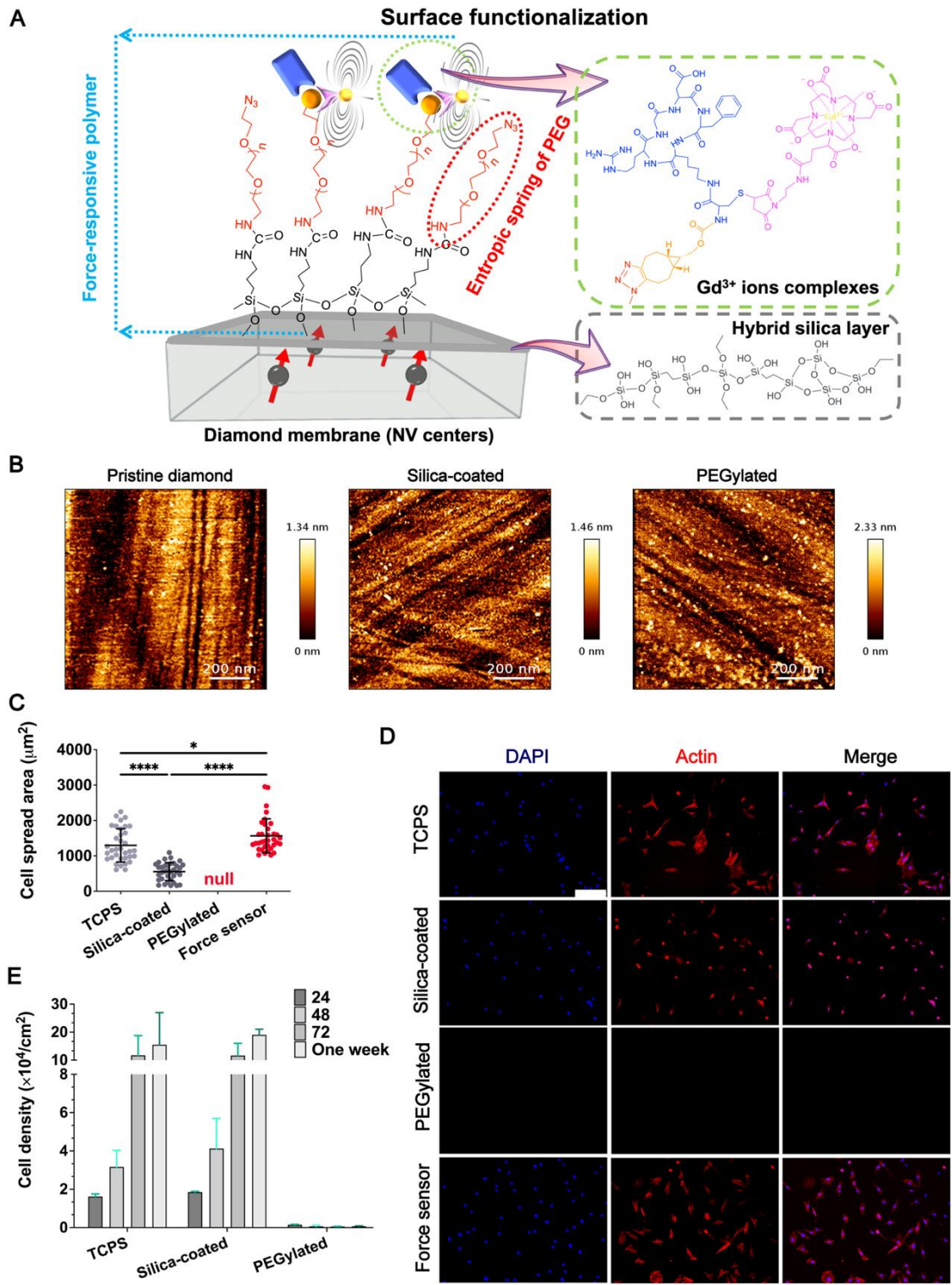

**Fig. 1. Diamond-based quantum sensing platform.** (**A**) Schematic illustration of overall chemical functionalization architecture. (**B**) AFM characterization of the pristine diamond, silica-coated and PEGylated surfaces of the diamond sensor. (**C**) Cell spreading area (n=40, three technical replicates, p values were obtained by one-way ANOVA followed by Tukey's post hoc test, mean with standard deviation (S.D)) and (**D**) representative images of NIH 3T3 stained with the cytoskeleton (Phalloidin, red) and nuclei (DAPI, blue) after culturing for 16 hours. The scale bar is 100 μm. (**E**) Attachment density of cells on diamond surfaces with different functionalization architectures after incubation for 1 day, 2 days, 3 days, and 7 days (mean with standard deviation (S.D)).

The X-ray photoelectron spectroscopy (XPS) results confirmed the successful conjugation of the force-responsive polymers onto the diamond surface (fig. S4 and Table S1). The length of the immobilized PEG polymer was ca. 6 nm in ambient conditions as detected by ellipsometry on the silica surface, while the thickness of a hybrid silica layer was about 2.9 nm (fig. S5). The functional diamond surfaces were analyzed by atomic force microscopy (AFM) to confirm the homogeneity of the deposited hybrid silica layer (Rq = 332.1 pm) and the polymer coating (Rq = 739.6 pm), and the typical aggregates of polymer brushes (*60*) could also be observed (Fig. 1B, Fig. 3A and fig. S6). These results further confirmed the successful conjugation of customized force-responsive polymers on diamond surfaces. In addition, the immobilized amount of the 2,2',2''-(10-(4-((2-(3-((3-((4-(14-benzyl-11-(carboxymethyl)-5-(3-guanidinopropyl)-3,6,9,12,15-pentaoxo-1,4,7,10,13-pentaazacyclopentadecan-2-yl)butyl)amino)-2-(((((1R,8S,9s)-bicyclo[6.1.0]non-4-yn-9-yl)methoxy)carbonyl)amino)-3-oxopropyl)thio)-2,5-dioxopyrrolidin-1-yl)ethyl)amino)-1-carboxy-4-oxobutyl)-1,4,7,10-tetraazacyclododecane-1,4,7-triyl)triacetic acid (BCN-RGD-DOTA (BRD)) reached 708 ng/cm$^2$ as detected by quartz crystal microbalance (QCM) with dissipation (fig. S7A). This density was sufficient to support cell adhesion and focal adhesion formation (*61*).

The stability of the immobilized polymer on the diamond surface was directly examined in cellular environments. The introduced polymers well supported the adhesion of NIH 3T3 fibroblasts during the whole cell culture period (Fig. 1C and D), while preventing the non-specific adsorption of bovine serum albumin (fig. S7B). Meanwhile, no cells could adhere on the polymer coatings without RGD ligands for at least 7 days (Fig. 1E and fig. S8). These results demonstrated that the polymer coating only interacted with cells through integrin-RGD adhesion and could keep stable for at least several days in cell incubation conditions. This was actually consistent with our reference test with the same polymer brushes on silicon wafer, i.e., the thickness of the PEG-coated layers (containing hybrid silica layer) kept constant after 5 days of immersing in phosphate-buffered-saline (PBS) buffer (fig. S5).

**Validation of quantum-enhanced force sensing**

To demonstrate the sensing capability of our customized widefield quantum diamond microscope (detailed in the methods section), we firstly investigated the influence of ferritin (a kind of paramagnetic species (*62*) as shown in Fig. 2A) solution on the NV spin relaxation in constructed PEGylated sample (without RGD ligands and magnetic labels). As shown in Fig. 2, B and C, the AFM image clearly indicated that there has been a dense packaged layer of ferritin formed on top of the diamond surface (after immersing in 1 mg/ml aqueous ferritin solution for 2 hours without rinsing). In the corresponding $T_1$ mapping images and histogram (Fig. 2D, left and middle panel, Fig. 2E), we found that the ferritin-deposited diamond surface showed a significantly shorter $T_1$ value (~ tens of µs), i.e., almost one order of magnitude lower than that in PBS cases (~ hundreds of µs). These findings indicated that the presence of ferritin molecules (or gadolinium ions as shown in fig. S9) significantly affects the NV spin relaxation as revealed by $T_1$ values (*62*). Interestingly, we have actually found the $T_1$ value could be recovered after gently washing with PBS (Fig. 2D, right panel, Fig. 2E), indicating the coated PEG chains successfully prevented the non-specific adsorption of proteins.

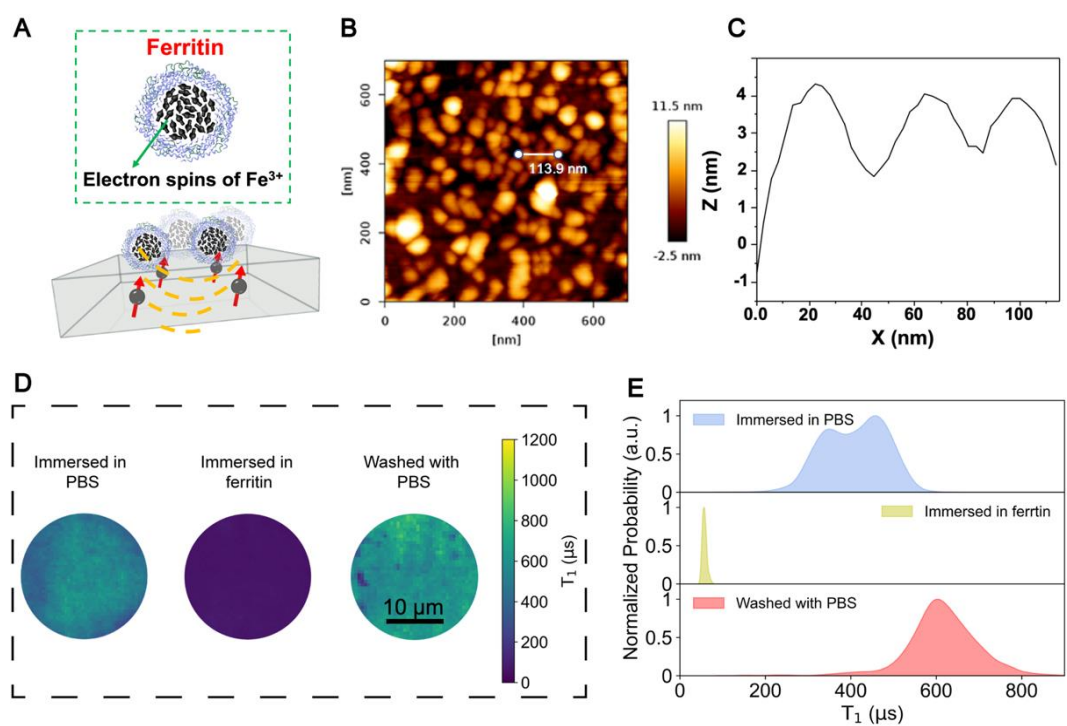

**Fig. 2. Sensing of magnetic labelled proteins using QDMTM. (A)** Top: a schematic illustration of ferritin, the black arrows depict the paramagnetic $Fe^{3+}$ ions. Bottom: schematic diagram of ferritin adsorption on PEGylated surface of diamond membrane. **(B)** AFM image of ferritin adsorbed on PEGylated surface and **(C)** the corresponding height profiles of ferritin adsorbed on PEGylated surface (positions marked in Fig. 2B). **(D)** $T_1$ mapping image of PEGylated diamond membrane immersed in PBS (left panel), 1.0 mg/ml ferritin (middle panel), and washed with PBS afterwards (right panel). **(E)** Corresponding histogram of $T_1$ mapping in Fig. 2D.

To validate the designed QDMTM, we firstly managed to alter the conformation (collapse-extended) of force-responsive polymers in model conditions, which mimics the polymer chain stretched by cellular forces. Specifically, it is well-known that the hydration (e.g., sample immersed in water) could extend the PEG chains while dehydration (e.g., sample exposed to air) would collapse the PEG chains (*63*). Therefore, the distance between the $Gd^{3+}$ magnetic labels and the NV centers can be modulated by placing the constructed diamond force sensor sample in different environments like water and air. The influence of solvents on the force sensor was confirmed by AFM, which exhibited the relatively homogenous microstructure with a surface roughness of 1198.0 pm for the decorated surfaces (water) (fig. S6). Fig. 3A and B show typical topographies of the polymer-modified diamond surface in air and water, disclosing the morphologies at the collapsed and extended states, respectively (the relative height increases by ~5.27 nm). The corresponding line profiles (positions marked in Fig. 3, A and B) were plotted in Fig. 3C, demonstrating a general tendency for molecular chain stretching caused by solvent effects. In addition, one end of PEG molecules is linked with $Gd^{3+}$ ion complexes and another end is attached to the diamond surface through a very thin "active" layer of hybrid silica. Therefore, extending or collapse of the polymer shell layer changes the NV-$Gd^{3+}$ distance and thus affects the measurement of NV spin relaxation time ($T_1$) (*43*). From the measured $T_1$ mapping of the force sensor diamond samples (Fig. 3, D and E), there have been much smaller $T_1$ values in air (collapsed status: shortened NV-$Gd^{3+}$ distance), compared with that measured in ultrapure water (extended status: prolonged NV-$Gd^{3+}$ distance). The same trend was found to maintain the same even after another cycle (fig. S10, A and B). Thus, the measured $T_1$ values were highly correlated with NV-$Gd^{3+}$ distance changes modulated by the PEG entropic spring, demonstrating the feasibility of our proposed QDMTM.

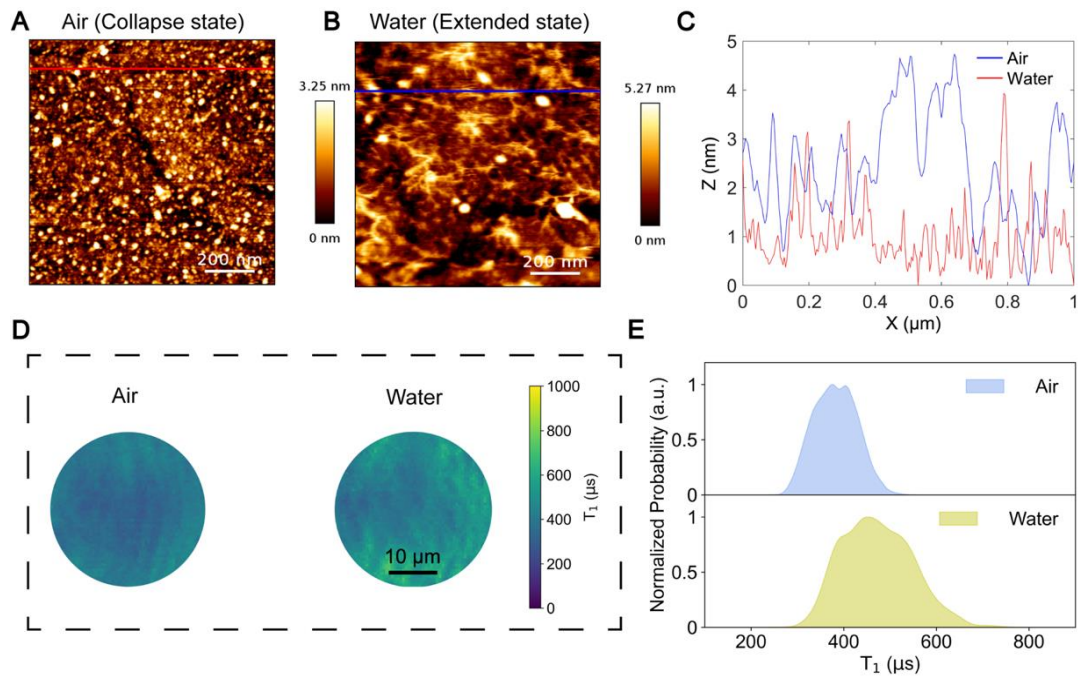

**Fig. 3. Validation of the designed QDMTM using the polymer conformational changes caused by solvent effects.** AFM images of constructed diamond force sensor in **(A)** air and **(B)** water,

respectively. **(C)** Line profile of images shown in Fig. 3A. **(D)** $T_1$ mapping of constructed diamond sensor in different states. **(E)** Corresponding histogram of $T_1$ mapping in Fig. 3D.

**Semi-quantitative mapping of cell adhesive forces**

By seeding the maturely adhered NIH 3T3 cells on diamond surface modified with force-responsive polymers, we started to demonstrate the detection of cellular adhesion forces using QDMTM. Cell adhesive force is transmitted to the environment through integrin-ligand interactions (*64*), and the force-responsive polymers were used to convert the mechanical input to the magnetic output. As illustrated in Scheme 1, the RGD end of the polymer was assumed to be recognized by integrins and dragged by the cellular traction force. The PEG entropic spring was stretched, and the $Gd^{3+}$ magnetic labels, located just next to the RGD ligands, were moved away from the diamond surface. This distance change could be quantitatively detected via the NV spin relaxometry, i.e., the larger force-induced conformation changes (of polymer) the longer the $T_1$ value is (of NV centers).

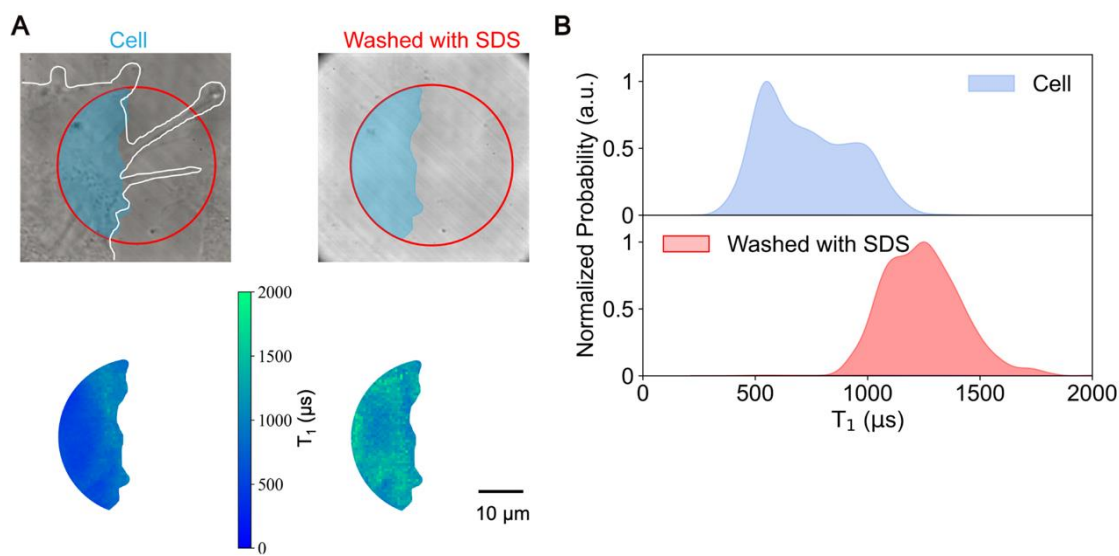

**Fig. 4. Validation of the designed QDMTM by living cells.** **(A)** Upper panel: transmitted light images of the selected regions in a typical NIH 3T3 cell grown on diamond force sensor before (left panel) and after (right panel) 1% SDS treatment; lower panel: $T_1$ mapping of the corresponding regions. **(B)** Corresponding histogram of $T_1$ mapping in lower panel of Fig. 4A. White dashed lines indicate the edge of the selected cell.

Based on the experimental results above and the theoretical analysis, we verified whether the variation in the distribution of $T_1$ was correlated with cellular regions. The adhered cells were detached by the treatment with 1% sodium dodecyl sulphate (SDS), and the $T_1$ value of the cell adhered region was recovered to the original level after cell removal (Fig. 4, A and B). This result confirmed that $T_1$ distribution was influenced by the adherent cells.

Meanwhile, as shown in Fig. 5A to C (i.e., a few typical examples), the relative $T_1$ changes (with respect to that measured in water as shown in Fig. 3) were larger in the peripheral area of the cells and sharped near the cell edges, where the pseudopodia and focal adhesions were enriched (fig. S11). This was consistent with the fact that the actomyosin stress fiber, the generator of cell traction force (*65*), mainly connected the two sides of the spread cells, thus the traction force concentrated at the cell edges, especially the edges of the polarized sides (*66*). For well-spread cells, the $T_1$ values in the polarized region of the cell (Fig. 5A, marked i) are greater than the region of the bridged edges of the cell (Fig. 5A, marked ii) and $T_1$ values in region ii are similar to those in region iii (Fig. 5D). Meanwhile, the relative $T_1$ changes in the region of the well-spread cells (Fig. 5, A and D) were larger than that of the less-spread cells (Fig. 5, B and D), indicating the larger adhesive force generated in the well-spread cells. This result matched the previous conclusion that the cell adhesive force was positively related to the cell spread area on the rigid or elastic substrates (*67*). Compared with single-cell measurements (Fig. 5, A and B), less $T_1$ changes were found in measurements with multiple cells (Fig. 5C, marked iv). This was probably due to the cell-cell interaction decreased the integrin-based adhesive forces, as part of the actomyosin-integrin linkage was disassembled to build the cell-cell adhesion (*68, 69*). The adhesive region of the less spread cells, which were contacted with neighbour cells (cell-cell contact), showed the lowest $T_1$ value (Fig. 5D), suggesting the smallest force of integrin adhesion. These experiments further validated the capability of QDMTM in measuring cell adhesive forces.

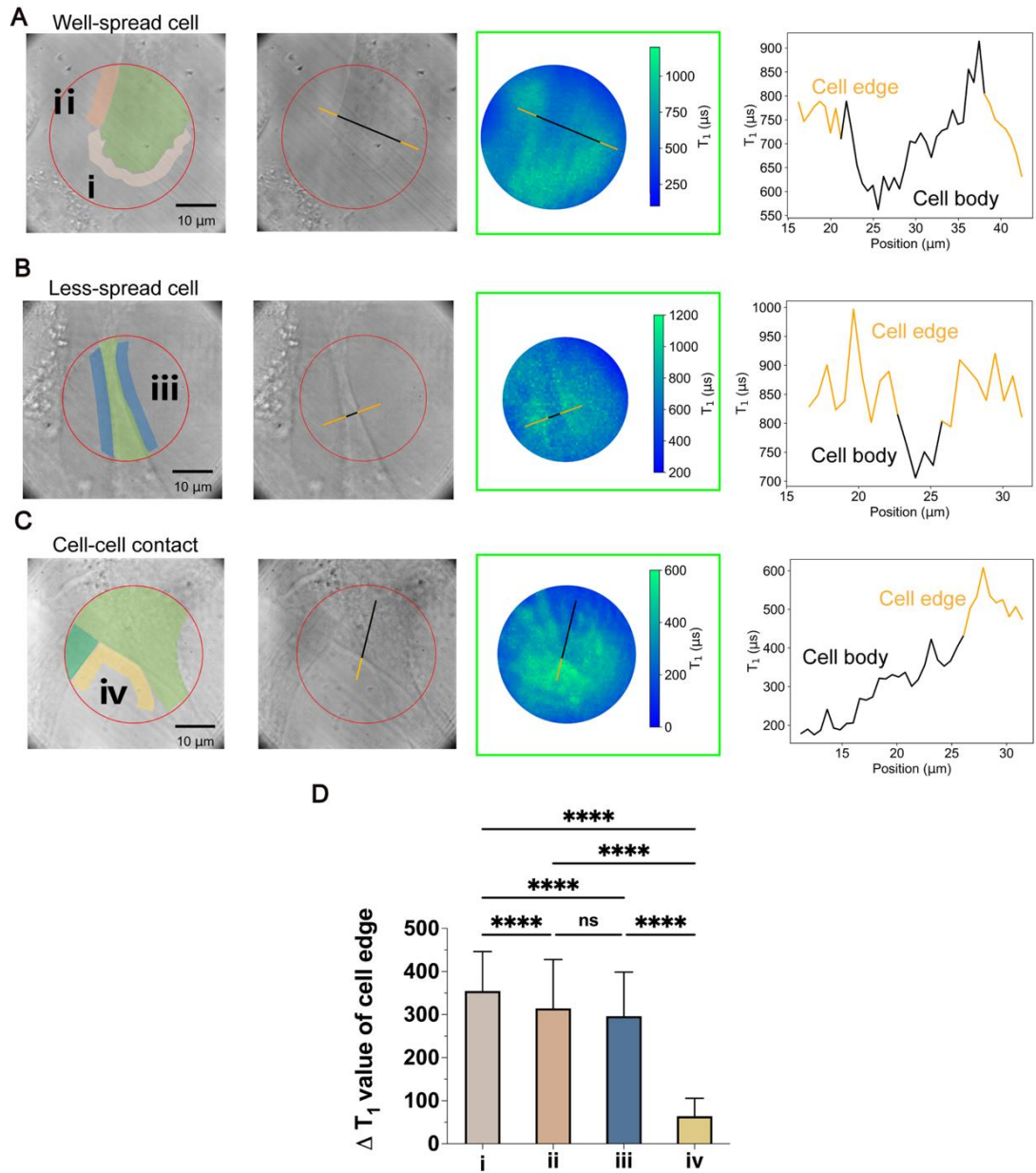

**Fig. 5. Demonstrated detection of the traction forces of the adhered cells.** Three typical cells, namely the **(A)** well-spread cell, **(B)** less-spread cell and **(C)** cell-cell contacted cells were chosen to demonstrate the measurements: the first two columns show the transmitted light images (left) and corresponding marked images (right panel, green represents the cell body while other colors represent the selected edges of cells); the third column shows the $T_1$ mapping and the last column shows the line profile, as-marked in 2$^{nd}$ and 3$^{rd}$ column, of $T_1$ value across the cell body. **(D)** $T_1$ values changes in selected areas of cells (denoted as i, ii, iii and iv as shown in Fig. 5, A to C, p values were obtained by one-way ANOVA followed by Tukey's post hoc test, mean with standard deviation (S.D). The detailed calculation of $\Delta T_1$ was described in the experimental section).

**Quantitative mapping of cell adhesive forces**

The force exerted on the QDMTM can be further quantified, as the cell adhesive forces could be revealed by relative $T_1$ changes (Fig. 5). Based on the model built above for measuring the cellular force (Fig. 6A), the distance between $Gd^{3+}$ molecules and NV centers plays a key role in determining $T_1$ values, and the quantitative relationship between them can be derived from Monte Carlo numerical simulation (more details shown in Methods). Meanwhile, PEG extending and collapse under cellular forces is unregulated, but the vertical distance between Gd and NV has the most significant effect on $T_1$, so here we consider the change of PEG length in the vertical direction. By combining the extended Worm-Like Chain (WLC) model, the relationships between cellular traction force and $T_1$ can be obtained as shown in Fig. 6B. According to the relationships, we have reconstructed the cellular force of the cell exerted on the PEG molecules, and the $T_1$ map can be converted into PEG extension map (Fig. 6C) and cellular force map (Fig. 6D). It is worth noting that the effective measurement range of cellular force is corresponding to the $T_1$ values between 391 μs and 795 μs (Fig. 6B). The $T_1$ reference value of 410 μs was obtained according to Fig. 3E. Thus, $T_1$ below the reference value is considered a force-responsive polymer being compressed by the cell and conversely extended. However, only the extension of the polymer fits the WLC model, the compression of the polymer can subsequently be investigated with a suitable model.

PEG length is an essential parameter both in experiments and simulations. PEG length influences the force-extension curve during the extended process. They determine the sensitivity and dynamic range of the sensor. In our experiment, the PEG (Mw: ~1000 g/mol) full contour length is 6.20 nm ~ 7.98 nm. It is more sensitive to the force range from several pN to around 30 pN (more details shown in Supplementary Information simulation). Based on the simulation result, extensions of PEG are ranging from 3.5 nm to 5.5 nm, and the tension loaded on single PEG is around 10 pN, which agrees with the previous study (*19, 70*). This semi-quantitative model can guide choosing suitable PEG length to tune the 'most sensitive range' of our sensor, making it competent for sensing different force ranges, enlarging the application scenarios of this new development kit.

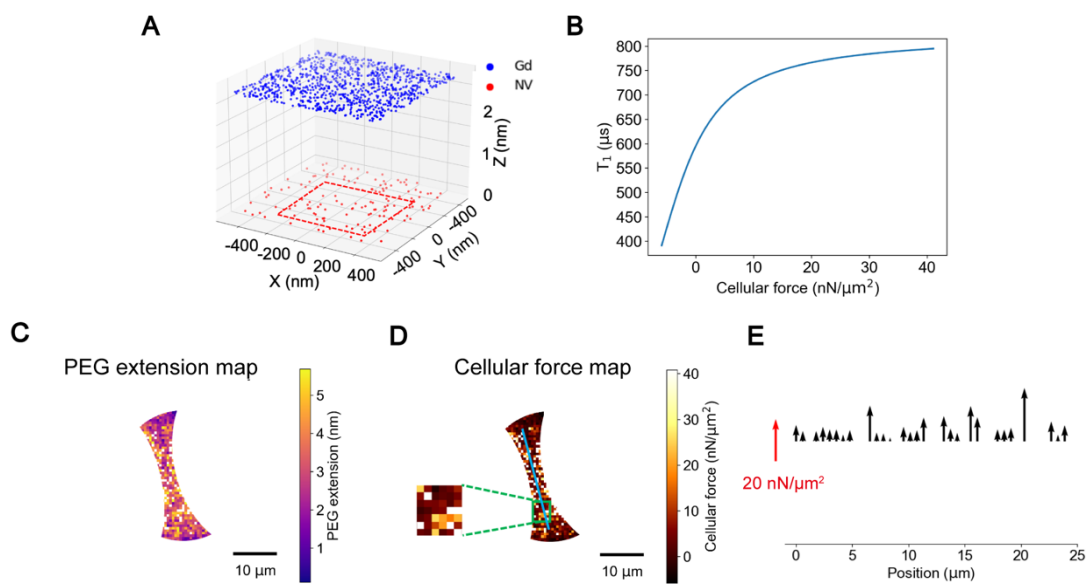

**Fig. 6. Simulation-assisted extraction of cellular forces from $T_1$ mapping.** (**A**) The schematics show the simplified model: a layer of NV centers is randomly distributed underneath the upper surface of diamond membrane with a given depth and fixed density; a layer of the $Gd^{3+}$ complexes attached to the PEG acts as a randomly fluctuating spin bath; the aforementioned two layers are separated by spring-like PEG molecules and their force-induced extension is described by Worm-Like Chain model. The red rectangular area represents the minimum sensing area adopted (600 nm × 600 nm). (**B**) The simulated relationship between cellular forces and $T_1$ value. The extracted (**C**) PEG extension map, and (**D**) cellular traction force map of cell body in one chosen cell (the same one in Fig. 5B). The insert is the enlarged view of selected rectangular area. (**E**) Force profile along the blue line drawn in Fig. 6D. The direction of the cellular force exerted on the PEG polymer is normal to the diamond surface.

## Discussions

Been contrary to traditional optical methods based on various fluorophores with a stochastic nature, our strategy relies on coupling the relevant mechanical signals to the spin states of the robust atomic defects in diamond, i.e., NV centers. It must be pointed out that the unprecedented sensitivity and precision of the proposed QDMTM is inherently guaranteed by the quantum nature of electronic spins of NV centers. This in conjunction with the fact that such approach is totally fluorophores label-free can in principle overcome various difficulties, such as photobleaching, limited sensitivity, and ambiguities in data interpretation.

The presented QDMTM successfully distinguishes the various mechanical states of the adhered cells, including the well-spread cell edges, the bridged cell edges, as well as the cell-cell contact region. The results of cell force in different regions are in line with the presented knowledge that the integrin adhesive force is concentrated in the cell polarized area and weakened by cell-cell interaction (*66, 68*). These applications indicate the effectiveness and precision of QDMTM for cell force measurement.

We have to admit that there is still room for improving QDMTM: 1) The stability of the optical system could be improved by introducing a fast auto-focusing module (*71*) as any sample drift might affect the $T_1$ measurements. 2) It is doable to adopt MW-free (*72*) and/or fixed-tau (*43*) schemes for speeding up the $T_1$ measurements. 3) The pressure of the cell bodies exerted onto the sensors can be calculated according to a suitable model, which may be helpful to understand the cell membrane tension.

Overall, this novel force-sensing tool, namely the QDMTM, will allow to enhance sensitivity as well as resolution in time and space in comparison to available traction force microscopy. Also quantum tension sensors may be reused after cleaning which will also enhance absolute precision of sensors for comparing different samples. It can fundamentally change the way how we study important issues like cell-cell or cell-material interactions, and hence bring impact to the field of biophysics and biomedical engineering. In addition, the data on the cellular forces transmitted through cell adhesions to be generated by this study are also expected to be useful in guiding and assisting the development of future theories on mechanosensing and mechanotransduction.

**Materials and Methods**

*Materials*

(3-isocyanatopropyl)trimethoxysilane (MeO)$_3$-Si-NCO), triethylamine (TEA),

tetraethyl orthosilicate (TEOS), 1,2-bis(triethoxysilyl)ethane (BTSE), $GdCl_3*6H_2O$, ethylenediaminetetraacetic acid disodium salt (EDTA-2Na), (1R,8S,9s)-Bicyclo[6.1.0]non-4-yn-9-ylmethyl N-succinimidyl carbonate (BCN-NHS) were purchased from Sigma-Aldrich (Shanghai, China). 2,2',2''-(10-(1-carboxy-4-((2-(2,5-dioxo-2,5-dihydro-1H-pyrrol-1-yl)ethyl)amino)-4-oxobutyl)-1,4,7,10 tetraazacyclododecane-1,4,7-triyl) triacetic acid (Maleimide-DOTA-GA) were purchased from Chematech (France). The peptide $NH_2$-cyclo(-Arg-Gly-Asp-D-Phe-Lys)-SH (CycloRGDfK) was synthesized by GL Biochem Ltd. (Shanghai, China). $N_3$-PEG-$NH_2$ (Mw: ~1000 g/mol) was synthesized by JenKem Technology (Beijing, China). All organic solvents are purchased from Acros (Germany) unless otherwise stated. The single crystalline diamond plates (2 mm × 2 mm× 0.03 mm, Applied Diamond Inc., Electronic Grade) were used to force sensor diamond by implanting $^{15}N^+$ ions into the diamond with a 5 keV per atom. The implanted nitrogen atoms have a mean depth of 5 ± 2 nm and are then annealed in a vacuum tube furnace to 800°C to form NV centers.

*Construction of sensing platform based on diamond membrane*

The single crystalline diamond plates (2 mm × 2 mm× 0.03 mm, Applied Diamond Inc., Electronic Grade; 3 mm × 3 mm × 0.25 mm, Element Six, Optical Grade) were chemically activated by freshly prepared piranha solution and a thin layer of hybrid silica was covalently modified on the cleaned pristine oxygen-terminated diamond, followed by Silane-PEG-$N_3$ was grafted on the silanization diamond surface and modification of BRD molecular by SPAAC reaction, finally, $Gd^{3+}$ ions were loaded and force sensor diamond can be obtained. To further regenerate the surface and promote recyclability, the force-responsive polymer modified diamond is easily cleaned by NaOH solutions and piranha washes. Additionally, the synthesis of BCN-RGD-DOTA molecule required two steps: first, CycloRGDfK reacted with Maleimide-DOTA-GA by click reaction; second, condensation of BCN-NHS reacted with the product of RGD-DOTA (Scheme S1 Supporting Information). Finally, it is modified to the surface by the SPAAC reaction of the azide group on PEG and $Gd^{3+}$ ions were loaded by chelation

(fig. S3). Detailed information on the synthetic force-responsive polymers and constructed diamond force sensor are provided in the Supporting Information.

*Characterization of constructed sensing platform*

The successful construction of quantum sensing platform for cellular force measurements, and characterization of the elemental composition, morphology, ligand density and thickness of functional surfaces by XPS, AFM, QCM and ellipsometry. As it was difficult to evaluate the thin layer coating on the diamond surface via standard ellipsometry, the polymer immobilization was analyzed on silicon model deposited with the silica layer (purchased from Ted Pella, Inc., 5 × 5 mm diced silicon wafer). Meanwhile, the silica QCM chips were used for the QCM test. Detailed information on the test parameters has been provided in the Supporting Information.

*Widefield quantum diamond microscopy*

This wide-field quantum diamond microscope mainly includes three subsystems – the optical system, the MW system and the control system. The optical system, based on a customized wide-field fluorescence microscope, provides efficient optical initialization and readout of NV centers in a large field of view. The 532 nm laser (Changchun New Industries, MGL-FN-532-1W) passing through an acousto-optic modulator (Gooch & Hoosego, 3250-220) is focused on the back-focal plane of a 100X oil objective with 1.50 NA (Olympus, UAPON100XOTIRF). The NV fluorescence is filtered with a long pass filter (Thorlabs, FELH0650) and imaged on a 512 × 512 air-cooled EMCCD (Teledyne Photometrics, Evolve 512 Delta) with an effective pixel size of 96 nm under the focal length of imaging lens at 300 mm (Thorlabs). A 561 nm long-pass dichroic mirror (Semrock, Di03-R561-t1-25 × 36) is for reflecting the excitation beam into the objective and collecting the fluorescence longer than 561 nm. In MW system, the microwave generated by a microwave source (ROHDE&SCHWARZ, SMBV100A) passing through a MW switch (Mini-Circuits, ZASWA-2-50DRA+) is amplified by a MW amplifier (Mini-Circuits, ZHL-16W-43-S+) and exerted on the NV centers by a customized Omega-shaped MW antenna, then transmitted to the terminator. The control system is made up of a computer and a pulse streamer (Swabian Instruments,

Pulse Streamer 8/2) for the data transfer and synchronizes the entire system.

*NV spin relaxometry measurements*

After being polarized by a green laser pulse, the electron spin of the NV centers will relax to thermal equilibrium state from polarized state which is named longitudinal relaxation. The longitudinal relaxation rate $\Gamma_1$ is principally dominated by spin-lattice interaction and fluctuating magnetic field generated by the nearby spin impurities, such as ferritin and $Gd^{3+}$ used in the experiments to modulate the relaxation rate.

In our sensing protocols, the NV center is polarized to $|0\rangle$ state by a 1 μs laser pulse with the power of 18 kW/cm$^2$ firstly. After 500 ns, a $\pi$-pulse of microwave is exerted to flip the NV center from $|0\rangle$ state to $|1\rangle$ state. Then after a waiting time $\tau$, another 1 μs laser pulse is applied to read out the spin state of the NV center and polarize the NV center. The above pulse sequences as a unit sequence will be repeated tens of thousands of times to obtain a bright enough fluorescent signal. Due to the camera's inability to switch on and off quickly, the camera remains in an exposure state throughout the tens of thousands of repetitions of the unit sequence with a specific waiting time $\tau$. To reduce the influence of background signal, we repeated the above unit sequence but turned off the microwave to perform a control measurement. Therefore, the NV center relaxed from the $|0\rangle$ state to the $|1\rangle$ state in this measurement. By performing element-wise division between the image matrices without and with microwave, the impact of the background signal will be minimized and the T$_1$ trace of the fluorescence intensity could be obtained. Finally, through fitting the T$_1$ fluorescence trace with the formula of $I(\tau) = a \cdot e^{-\left(\frac{\tau}{T}\right)^b} + c$, we can get the T$_1$ value.

*Seeding cells on the diamond-based sensing platform*

According to the cell force measurement, we immobilized force sensor diamond (2 mm × 2 mm × 0.03 mm, Applied Diamond Inc., Electronic Grade) onto microwave antennas with an omega structure (280 microns in diameter) by using PDMS cured at 60°C (Dow Corning Sylgard 184; monomer: crosslinker = 10: 1). Samples were sterilized for 1 hour with 70% ethanol. The density of 10$^4$/ml NIH 3T3 cells were cultured on the above-encapsulated diamond slides for 6 h and stability of cell adhesion was observed.

*Measurements of cell adhesion forces via NV spin relaxometry*

Cells were washed once with cell culture medium and twice with PBS before fixation with 4% paraformaldehyde at room temperature for 15 min. Force sensor samples with adherent mature cells were then washed three times with PBS. Thoroughly cleaned diamonds were immersed in PBS for $T_1$ testing at room temperature. For $\Delta T_1$ of Fig. 5D obtained by selecting the region $T_1$ (Fig. 5, A to C, marked i, ii, iii and iv) minus the reference value (obtained by Fig. 3E).

*Theoretical model for quantifying the relationship between cell forces and $T_1$ value*

A numerical simulation model is built to simulate the longitudinal relaxation time of the NV center under the magnetic disturbance of the $Gd^{3+}$ ions (More information about the numerical model please reference to Supporting Information). In this model, the location of the NV centers is randomly generated with an orientation randomly selected from four given directions of $[\bar{1}11]$, $[1\bar{1}1]$, $[11\bar{1}]$ and $[\bar{1}\bar{1}\bar{1}]$ by the Monte-Carlo Simulation (the bulk diamond is [100] cut). The density and depth of the NV centers are $1000/um^2$ and 5 nm (under the bulk diamond surface), respectively. The density of the $Gd^{3+}$ is set to $9000/um^2$ in simulation. The location of the $Gd^{3+}$ in XY-plane is also generated randomly by Monte-Carlo simulation.

For calculating the $T_1$ of a single NV center, we only take the $Gd^{3+}$ ions (with a same height) inside a circle with the diameter of 100 nm over the NV centers into consideration, because the NV centers outside the circle have no interactions with the NV centers. Then recover the $T_1$ fluorescence curve of the single NV centers based on the $T_1$ value obtained in the simulation. Sum all the $T_1$ fluorescence curves of the NV centers within the red dashed square of $600 \times 600$ $nm^2$ (Fig. 6A) which corresponds to the effective 6 × 6 binned pixel size on the sample plane under our setup configuration. Finally, the $T_1$ value of the NV centers under the interaction of the $Gd^{3+}$ ions could be obtained by fitting the summed $T_1$ fluorescence curve with a formula of $I(t) = A \cdot e^{-\left(\frac{t}{T}\right)^b} + c$. By scanning the height of the $Gd^{3+}$ ions from 0.3 to 6 nm in the numerical model, a relationship of the height of $Gd^{3+}$ ions and the $T_1$ values could be given, where the height of $Gd^{3+}$ ions represents the length of the PEG polymers. Finally, combining the Worm-Like Chain model (Refer to supporting information for details.) where the relationship of the force exerted on the PEG and extension of the PEG could be described, the relationship of the $T_1$ value and the force exerted on the PEG could be given.

**Acknowledgments**

**Funding:** Q.W. acknowledges the financial support from the National Natural Science Foundation of China (Grant T2222020) and the Sichuan Science and Technology Program (No. 2020YFH0034). Z.Q.C. acknowledges the financial support from the HKSAR Research Grants Council (RGC) Research Matching Grant Scheme (RMGS, No. 207300313); HKSAR Innovation and Technology Fund (ITF) through the Platform Projects of the Innovation and Technology Support Program (ITSP; No. ITS/293/19FP); HKU Seed Fund; and the Health@InnoHK program of the Innovation and Technology Commission of the Hong Kong SAR Government.

**Author Contribution:** F.X. and S.X.Z. contributed equally to this work. Z.Q.C., Q.W., J.W., and J.S. conceived the idea. F.X., S.X.Z., Y.H., and J.L. performed the experiments and analyzed data under the supervision of Q.W. and Z.Q.C.. L.J.M. performed the simulations. A.D. prepared the diamond membrane with shallow NV centers. F.X., S.X.Z., Q.W. and Z.Q.C. wrote the manuscript with input from all authors.

**Competing interests:** The authors declare that they have no competing interests.

**Data and materials availability:** All data needed to evaluate the conclusions in the paper are present in the paper and/or the Supplementary Materials.

# Supplementary Materials for

# Quantum-Enhanced Diamond Molecular Tension Microscopy for Quantifying Cellular Forces

Feng Xu[#], Shuxiang Zhang[#] *et al*.

* Corresponding author: zqchu@eee.hku.hk (Zhiqin Chu), wei@scu.edu.cn (Qiang Wei)

**This PDF file includes:**

Figs. S1 to S14
Table S1
Scheme S1
References (73 to 82)

**Synthesis of BRD (BCN-RGD-DOTA)**

Synthesis of BRD was carried out in two-step reaction with high conversion rate, as shown in scheme S1. The 3.0 ml of dry dimethylformamide (DMF) was taken and added into 10 ml flask under argon flux to dissolve cyclo(-Arg-Gly-Asp-D-Phe-Lys) (CycloRGDfK, 3 mg, 0.004 mmol, Mw: 706.32 g/mol). Additional 1.0 ml DMF solution was taken to dissolve 2,2',2"-(10-(1-carboxy-4-((2-(2,5-dioxo-2,5-dihydro-1H-pyrrol-1-yl)ethyl)amino)-4-oxobutyl)-1,4,7,10 tetraazacyclododecane-1,4,7-triyl) triacetic acid (Maleimide-DOTA-GA, 3.1 mg, 0.0052 mmol, Mw: 598.26 g/mol, 1.25 eqv. to thiol groups of CycloRGDfK). The Maleimide-DOTA-GA solution was slowly added into the CycloRGDfK solution by dropwise. The mixture was stirred at room temperature for 24 hours. Further purification was not required, and it was directly used in the next step. The m/z of RGD-DOTA was detected by LC-MS. $M^+$: m/z 1304.58; found m/z 1305.6 ($M^+$ + H); m/z 435.9 ($M^+$ + 3H)/3; m/z 653.3 ($M^+$ + 3H)/3. The CycloRGDfK peak was undetected, demonstrating the high reaction efficiency.

Next, the (1R,8S,9s)-Bicyclo[6.1.0]non-4-yn-9-ylmethyl N-succinimidyl carbonate (BCN-NHS, 0.88 mg, 0.003mmol, 0.8 eqv to amine of RGD-DOTA) was dissolved in 1.0 ml of DMF solution and added to the obtained RGD-DOTA mixture, followed by stirring at room temperature for 24 hours. After the reaction, the BCN-RGD-DOTA solution was obtained. The m/z of the products were analyzed by MADTI-TOF-MS. $M^+$: m/z 1480.67; found m/z 1481.1066 ($M^+$ + H). The RGD-DOTA and BCN-RGD-DOTA peaks were detected, indicating successful synthesis of BCN-RGD-DOTA.

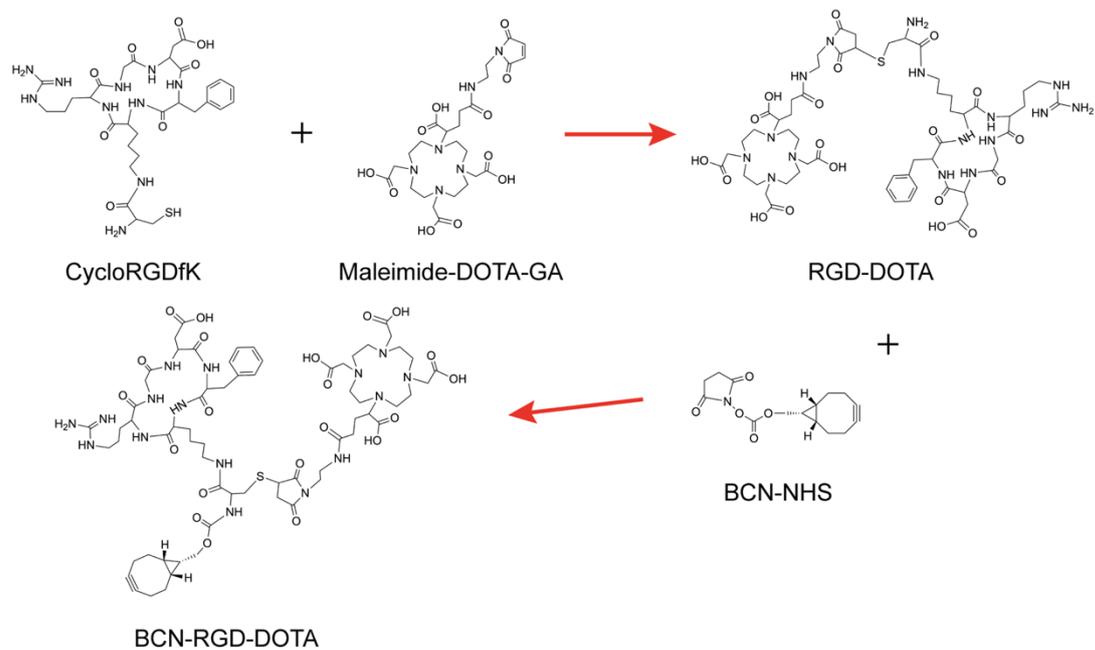

**Scheme S1.**

General procedure for the synthesis of the BCN-RGD-DOTA (BRD).

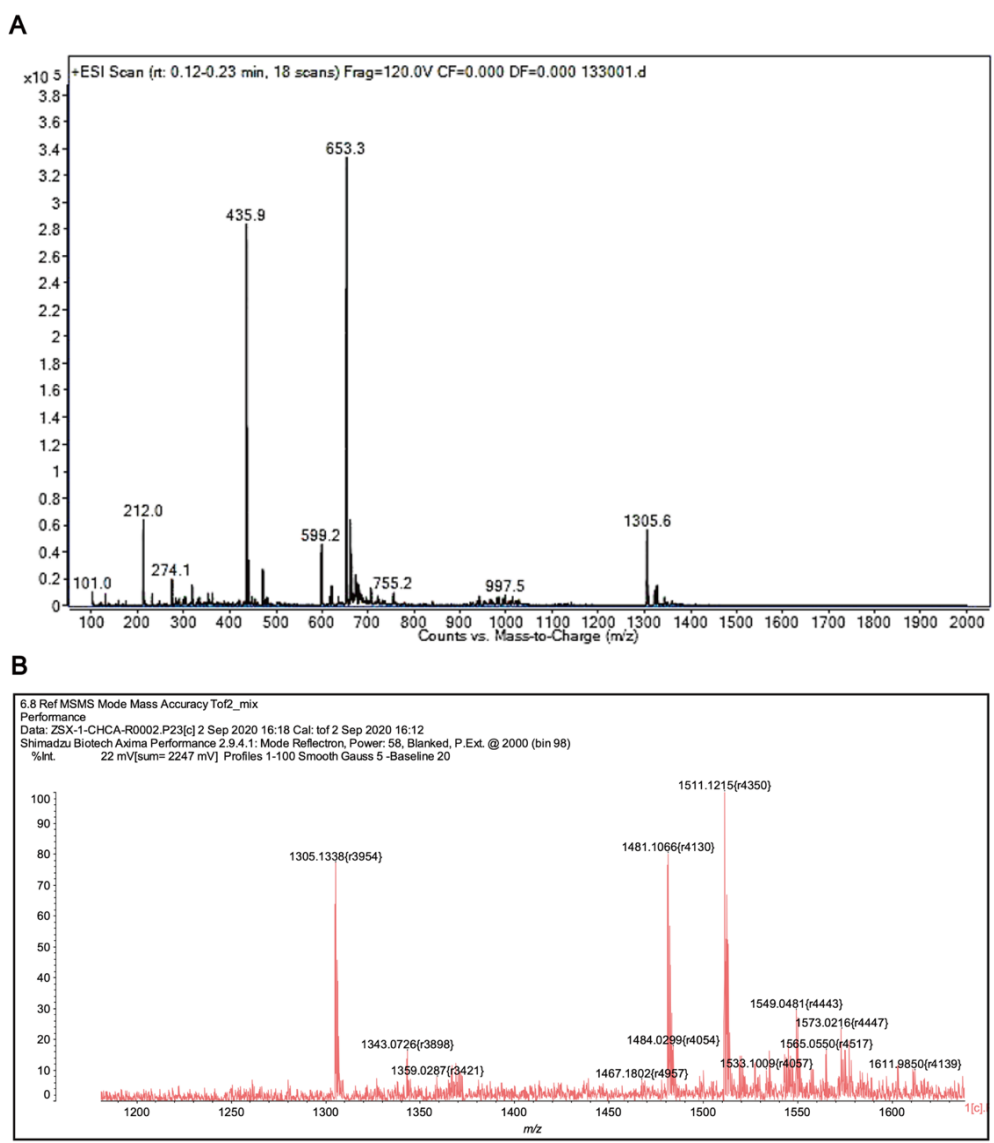

**Fig. S1.**

**(A)** LC-MS spectrum of the crude product of RGD-DOTA. **(B)** MADTI-TOF-MS spectrum of the crude product of BCN-RGD-DOTA (BRD).

**Synthesis of Silane-PEG-N₃**

The 1 ml toluene was added into the flask under argon flux to dissolve N$_3$-PEG-NH$_2$ (10.0 mg, 0.01 mmol, Mw: ~1000 g/mol). The 4.0 μl of (3-isocyanatopropyl)trimethoxysilane ((MeO)$_3$-Si-NCO, 2 eqv. to amino groups) and 1.5 μl of triethylamine (TEA) were added into reaction flask under argon flux as well. The argon flux was maintained for 30 min, and then the reaction solution was kept stirred at 60°C for 24 hours. The products were purified by precipitation from n-hexane to obtain a viscous yellowish liquid (Silane-PEG-N$_3$). ¹H NMR (DMSO-d6), δ ppm: 0.4-0.6 (Si-C**H**$_2$, 2H), 1.1-1.2 (NH-CH$_2$-C**H**$_2$-CH$_2$, 2H), 2.8-3.0 (N$_3$-C**H**$_2$, 2H), 3.0-3.2 (N$_3$-CH$_2$-C**H**$_2$, NH-C**H**$_2$-CH$_2$, 4H), 3.45-3.65 (PEG backbone, O-C**H**$_3$, C**H**$_2$-C**H**$_2$-NH).

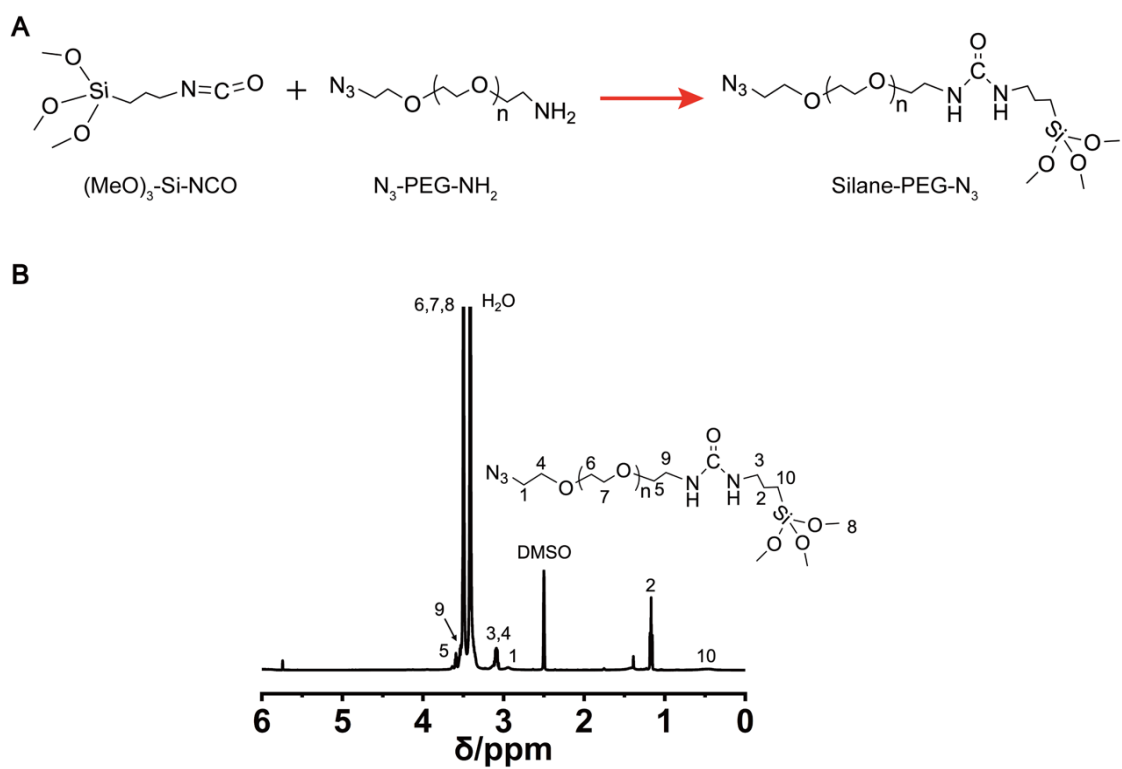

**Fig. S2.**

(**A**) General procedure for the synthesis of the Silane-PEG-N$_3$. (**B**) ¹H NMR spectrum of Silane-PEG-N$_3$ in DMSO-d6.

**Constructing force-responsive polymers on diamond surface**

Single-crystalline diamond slides were sonicated in acetone and isopropanol for 5 min in each and dried with nitrogen. The diamond slides were cleaned and chemically activated by freshly prepared piranha solution ($H_2SO_4/H_2O_2$=7:3) at 90°C for 1 hour, followed by thoroughly rinsing with ultrapure water and ethanol as well as drying with nitrogen (the samples were named as Pristine Diamond).

The 20 µl tetraethyl orthosilicate (TEOS) was added to a mixture of ethanol (2850 µl), ultrapure water (150 µl) and hydrochloric acid (10 µl) by dropwise for 1 hour. Then 10 µl of 1,2-bis(triethoxysilyl)ethane (BTSE) was added by dropwise as well, and the hydrolysis reaction was continued for another 1 hour. Afterwards, the cleaned Pristine Diamond was placed into this solution for 6 hours. After the reaction, the diamond slides were cleaned with ethanol and dried with nitrogen, obtaining hybrid silica-modified diamond surfaces (the samples were named as Silica-coated).

Then, the Silica-coated diamond slides were placed in a flask and immersed in a solution of 10 mg/ml Silane-PEG-$N_3$ dissolved in toluene. The reaction solution was kept at 55°C for 24 hours. After the reaction, the cleaned slides were kept under vacuum at 80°C for 1 hour and in an incubator at 60°C overnight. Subsequently, it was immersed in an anhydrous ethanol solution for 12 hours to quench the unreacted isocyanate groups on the slide surfaces and obtained PEGylated diamond surfaces (the samples were named as PEGylated)

The PEGylated surfaces were immersed in 1.0 ml of the BRD solution for 24 hours. Afterwards, the slides were removed and washed with dimethylformamide and ethanol, followed by drying with nitrogen to obtain biofunctionalized diamond surfaces. Finally, the gadolinium ions were loaded by immersing the BRD-modified diamond slides into 0.5 mg/ml $GdCl_3*6H_2O$ water solution for 2 hours and washed with ethylenediaminetetraacetic acid disodium salt (EDTA-2Na, 0.4 mg/ml) for 1 hour. after the chelation, the slides were cleaned with water and ethanol, and dried with nitrogen to achieve force-responsive polymers modified diamond surfaces (the samples were named as Force sensor).

Silicon wafers were used for similar modification processes, named Si, Si-Silica-coated, Si-PEGylated, and Si-Force sensor.

**Reuse of the diamond slides**

The Force-sensor can be simply removed by NaOH and piranha solution, providing an easily recyclable NV quantum sensor. Briefly, the functionalized diamond slides were firstly immersed in 1 M NaOH solution at 80°C for 12 hours, and then in piranha at 90°C for 1 hour. The corroded slides were extensively rinsed with ultrapure water and sonicated with acetone and 2-isopropanol for 5 min in each and dried with nitrogen. Besides, the slides can be also soaked in a 1:1:1 mixture of nitric acid, perchloric acid and sulphuric acid at boiling temperature (*73*).

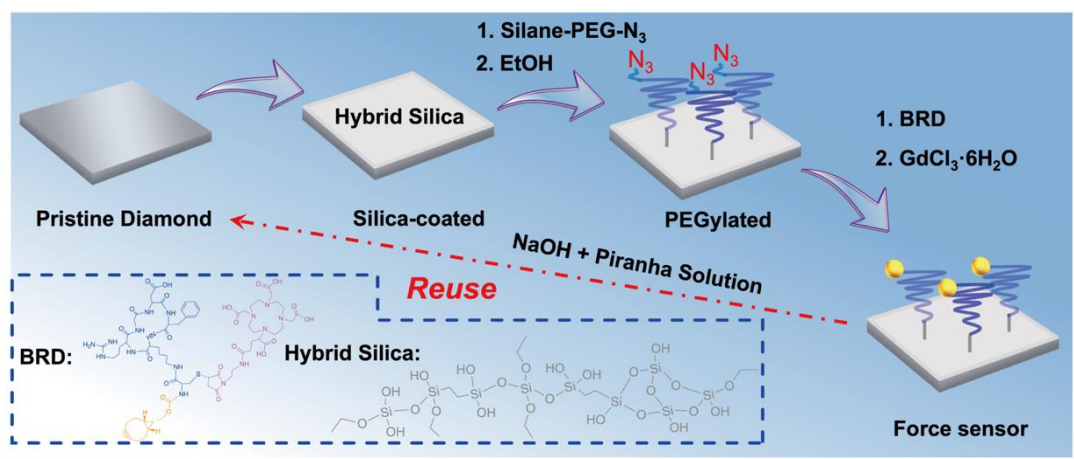

**Fig. S3.**

Schematic illustration of the functionalization process. The ultrathin hybrid silica layer was fabricated by BTSE and TEOS. The cell adhesive molecule (BCN-RGD-DOTA), consisting of an adhesion peptide (CycloRGDfK, blue in BRD) and a chelator (DOTA, pink in BRD), is immobilized to the azide terminal of the PEG polymer by the BCN-based (orange in BRD) SPAAC reaction.

**Surface characterization**

At each step of the functionalization procedure, we characterized the elemental composition, morphology and thickness of the functional surfaces by XPS, AFM, QCM and ellipsometry. Silicon wafer is utilized as model surface, instead of diamond slides, because it adapts to various surface characterization techniques such as ellipsometry, QCM and XPS (*74*).

**X-ray photoelectron (XPS) spectroscopy**

The surface elemental tests were carried out on silicon wafers with different coating steps by XPS (K-Alpha XPS, Thermo Scientific). The XPS measurements were performed on a Kratos system with $4 \times 10^{-10}$ mbar base pressure, sample neutralization applying low energy electrons, hybrid mode, take off angle of electrons (0°), pass energy (160 eV), and excitation of photoelectrons by monochromatic $Al_{k\alpha}$ radiation (hv = 1486.6 eV) at 300W (15 kV Å~ 20 mA). The detected region was elliptically shaped (300 µm × 700 µm for main axes).

XPS analysis was used to quantitatively determine the chemical composition of the surface of the silicon wafers modified with hybrid silica and PEG. Compared to the unmodified Si, the intensity of the C-O (286.8 eV) and Si-O (103.6 eV) peaks on the Si-PEGylated and Si-Silica-coated surfaces were obviously increased (fig. S4B), wherein C-O and Si-O were affiliated to the backbone of PEG and silane, respectively.

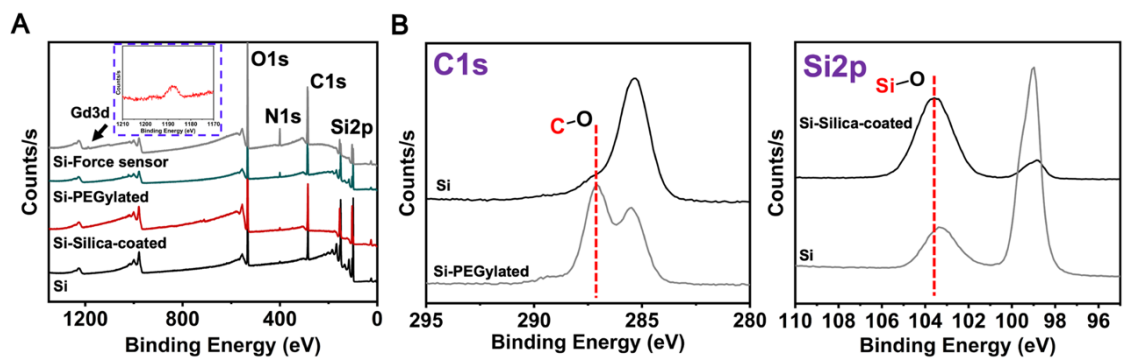

**Fig. S4.**

**(A)** XPS survey spectrum of C1s, N1s, O1s and Si2p signals after each functionalization step (insert represents Gd3d signal spectrum). **(B)** High resolution spectra of C1s and Si2p spectrum of Si, Si-Silica-coated and Si-PEGylated

Table S1 showed the relevant atomic compositions of different functionalized surfaces. After silica deposition, the content of O changed notably. The PEG grafting increased the N signal from 1.86 to 2.99% and the C/Si ratio was shifted to 1.43. After immobilizing the BRD and loading $Gd^{3+}$, the N signal further increased from 2.99 to 6.2%, the Gd signal appeared, and the C/Si ratio was shifted to 2.91, confirming the presence of BRD and $Gd^{3+}$.

In addition, after the chelation, EDTA-2Na was used to clean the free $Gd^{3+}$ trapping in the PEG chains. The stability constants of EDTA-2Na with $Gd^{3+}$ ions were lower than that of DOTA, thus, the chelated $Gd^{3+}$ was stable during purification (*75*).

**Table S1.**
XPS elemental surface composition of Si, Si-Silica-coated, Si-PEGylated, and Si-Force sensor.

| Samples | Elements (%) | | | | | C/Si |
|---|---|---|---|---|---|---|
| | [Si] | [C] | [N] | [O] | [Gd] | |
| Si | 42.95 | 23.42 | 1.8 | 31.83 | / | 0.55 |
| Si-Silica-coated | 24.8 | 29.54 | 1.86 | 43.8 | / | 1.19 |
| Si-PEGylated | 27.82 | 39.9 | 2.99 | 29.29 | / | 1.43 |
| Si-Force sensor | 16.44 | 47.89 | 6.2 | 29.65 | 0.11 | 2.91 |

**Ellipsometry**

Silicon wafers with different steps of coatings were tested by ellipsometry, which was performed in the spectrum range of 380 to 1050 nm at the incidence of 70°, with an ellipsometer (SENpro, SENTECH Instruments GmbH, Germany). Each data point resulted from an average of at least 3 measurements, and the obtained sensor grams were fitted with a four-layer model (Si, SiO$_2$, organic layer, and air) using the analysis software SpectraRay/3. The model layer of 'silicon VIS+NIR' was used as substrate with n = 3.817 and k = 0.01576. The thickness of the PEG and hybrid silica layers was measured without BRD decoration. The layers were set as Cauchy layer and assumed to be constant (T = 2.0 nm, N0 = 1.56, and N1 = 104.5). The layers were further fitted by the Cauchy model.

**Stability test**

For the coating stability test, we soaked the Si-PEGylated in PBS solution for 5 days and then rinsed them three times with ultrapure water before drying with nitrogen gas.

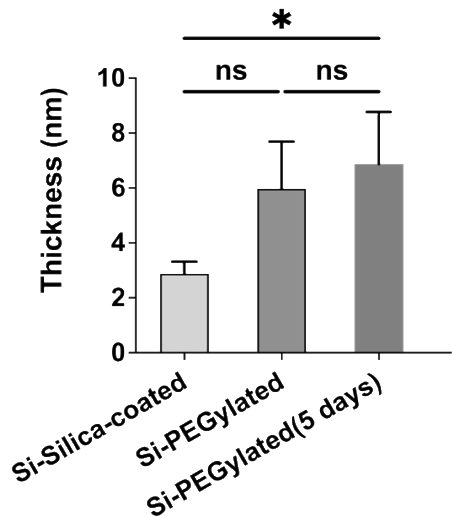

**Fig. S5.**

Ellipsometry results showing the thickness of Si-Silica-coated and Si-PEGylated surfaces on silicon wafers before and after immersion in PBS solution for 5 days at room temperature (p values were obtained by one-way ANOVA followed by Tukey's post hoc test, mean with standard deviation (S.D)).

**Atomic force microscopy (AFM)**

The surface morphology of the modified diamonds was recorded by NanoWizard 4 XP scanning probe microscopy (SPM) system (Bruker, USA) in the air and water under ambient conditions. The commercially available AFM Probe (TESP-V2) with a spring constant of ~ 37 N/m and resonance frequency of ~ 320 kHz was used in Tapping Mode, and the scanning rate was set at 0.8 Hz. SCANANSYST-AIR probes with a spring constant of 0.4N/m (Bruker, USA) were used in Quantitative Imaging (QI) Mode. The average surface roughness of the tested surfaces was analyzed by JPK Data Processing software and calculated from AFM images (1 × 1 µm² for functionalization for diamonds and 0.6 × 0.6 µm² for ferritin absorb on the PEGylated).

As confirmed by atomic force microscopy (AFM), we were able to deposit a uniform hybrid silica layer (Silica-coated: root mean roughness $R_q$ = 332.1 pm) on the oxygen-terminated diamond surface (Pristine Diamond: $R_q$ = 304.1 pm). The roughness slightly increased after PEG immobilization (PEGylated: $R_q$ = 528.7 pm). The final surface roughness $R_q$ = 739.6 pm can be obtained after the modification of force-responsive polymer (Force sensor).

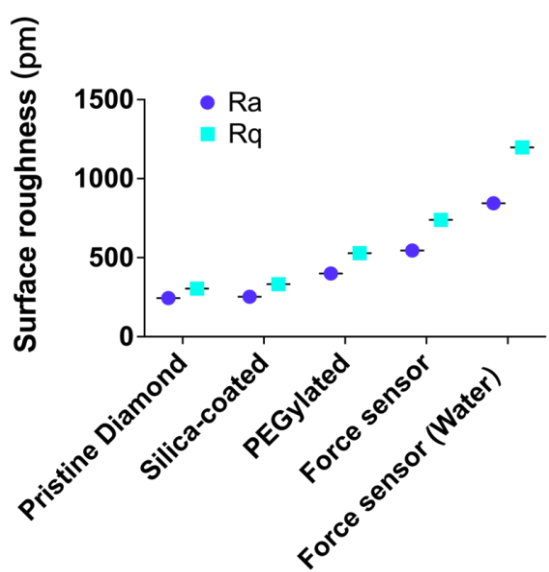

**Fig. S6.**

AFM characterization of the roughness of the functional diamond surfaces after each functionalization step.

**Quartz crystal microbalance**

Quartz crystal microbalance (QCM) with dissipation (Q-Sense E1, Sweden) was used to study the online coating of BRD and the BSA adsorption on the surfaces with different modifications. QCM with dissipation allows the monitoring of changes in resonance frequency ($\Delta f$) and dissipation ($\Delta D$) of a piezoelectric quartz crystal as a function of time. The f and D were recorded at the fundamental frequency (4.95 MHz) and its 3rd, 5th, 7th, 9th, 11th, and 13th overtones. Only the 3rd overtone was shown in the sensor grams.

The whole measurement was performed at 25 °C. The Sauerbrey equation was used to calculate the mass of the adsorbates [$\Delta m = C \times \Delta f$, where $\Delta m$ is the change in mass, C is the mass sensitivity constant of the quartz crystal (-17.7 ng·cm$^{-2}$·Hz$^{-1}$), and $\Delta f$ is the overtone-normalized frequency change] as the $\Delta D$ values were low.

The PEG-modified silica QCM chips (LOT-Quantum Design GmbH, Darmstadt, Germany) were fabricated via the protocol above. For monitoring the online coating of BRD molecular, the cleaned PEG-coated chips were inserted into flow chamber (QFM 401, QSense, Sweden, internal volume of 40 μl) and incubated in DMF/$H_2O$ (1:9 v/v) with a flow rate of 0.1 ml/min. After baseline equilibration, a solution of BRD (Figure S7a, 1 mg/ml in DMF/$H_2O$ (1:9 v/v)) was pumped into the flow chamber at the same rate. After 1 hour of online incubation, the flow chamber was alternately rinsed with Milli-Q water, aqueous solution of deconex 1% (w/w, Borer Chemie AG, Switzerland), and Milli-Q water.

The protein adsorption was measured similarly. The coated sensors were inserted into the titanium flow chamber (QFM 401, Q-Sense, Sweden, internal volume of 40 μl) and incubated in PBS buffer. After baseline equilibration, PBS buffer was pumped into the flow chamber for 10 min, and then the protein solution 1 mg/mL BSA solution was pumped into the flow chamber, followed by washing with PBS. Fig. S7B shows the BSA absorption on the PEG and complete force-responsive polymers modified-silica QCM chip. The coatings were efficient enough to resist the protein

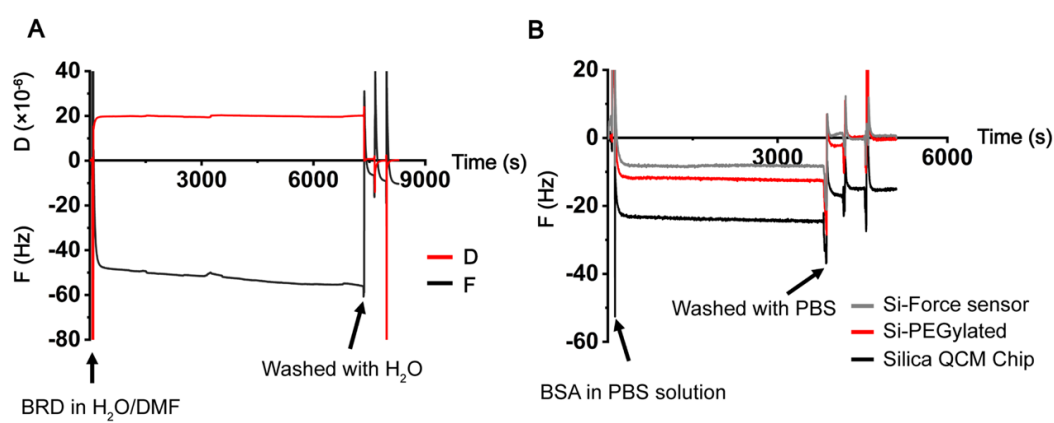

**Fig. S7.**

**(A)** QCM frequency (F) and dissipation (D) shift as a function of time during binding of BCN-RGD-DOTA (BRD) on PEG-coated gold QCM chip (contain hybrid silica layer) in DMF/$H_2O$ (v/v: 1:9). **(B)** QCM frequency shift of the adsorption of bovine serum albumin (BSA) on silica QCM Chip, Si-PEGylated, Si-Force sensor.

After 3 days of cell culture with NIH 3T3 fibroblasts, the tissue culture polystyrene (TCPS) and Silica-coated diamond surfaces were covered with well-spread cells. Whereas almost no cells adhered on PEGylated surfaces (Fig. S8).

Siloxane materials are reported to be not hydrothermally stable, herein, BTSE not only produces more Si-OH but also increases the stability of the silica layer (*76, 77*). Besides, the hybrid silica can be further modified without decreasing the sensitivity of the measurement. These data demonstrated that the introduction of an "active" hybrid silica layer on the diamond surface is imperative to stabilize the force-responsive polymers.

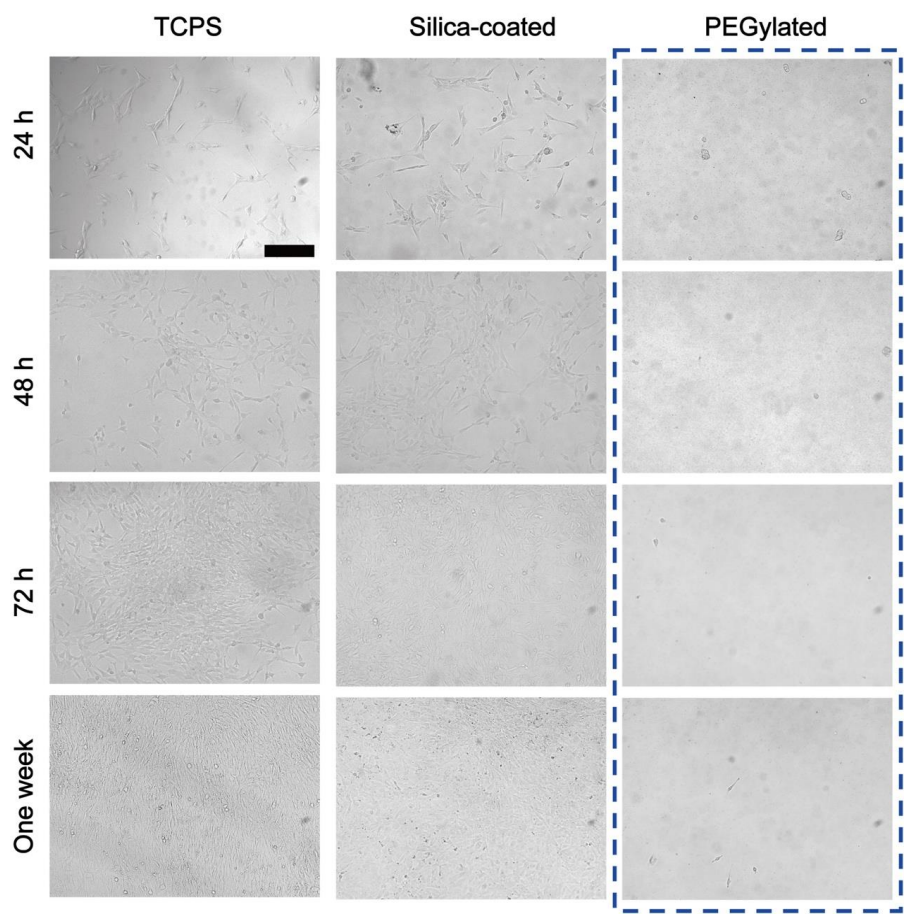

**Fig. S8.**

Optical microscopy images of NIH 3T3 cells adhered on the Silica-coated, PEGylated functional diamond surfaces after 1 day, 2 days, 3 days, and 7 days of cell culture (Scale bar indicates 100 μm).

When the diamond slides were immersed in the 1 mM GdCl$_3$ solution, the T$_1$ decreased 13 times compared with it in the pure water. We blocked part of the diamond with polydimethylsiloxane (PDMS). The PDMS decreased the T$_1$ value of the slide to 68% in pure water, which may attribute to the impurity of the metal catalyst for PDMS synthesis (*78*). After immersing in the GdCl$_3$ solution, the T$_1$ value of the PDMS blocked region was 4 times higher than the unblocked region because the PDMS reduced the diffusion of Gd$^{3+}$ to the diamond surface.

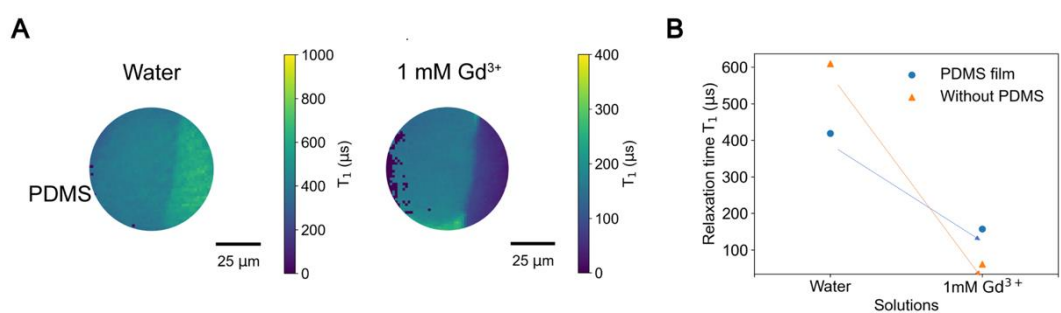

**Fig. S9.**

(**A**) T$_1$ mapping of pure diamond membrane (without surface modifications) in the presence of ultrapure water and 1 mM Gd$^{3+}$ solution, respectively, part of which was covered by PDMS (40X air objective). (**B**) Corresponding mean values of T$_1$ mapping in Fig. S9A.

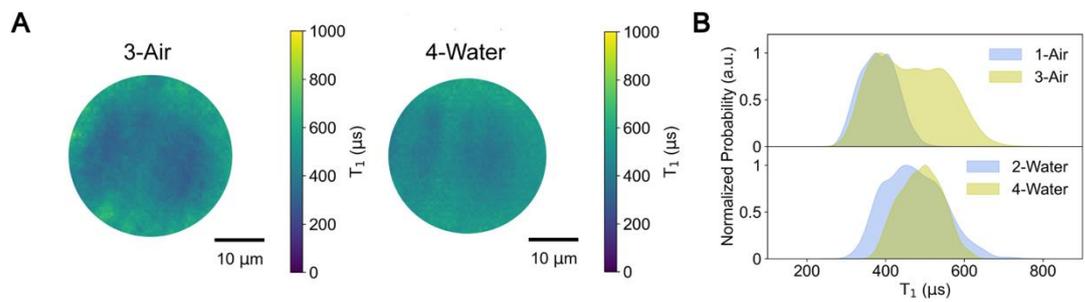

**Fig. S10.**

**(A)** $T_1$ measurements of the force sensor in air and ultrapure water environments after another cycle, respectively. **(B)** The histogram of $T_1$ values within the $T_1$ mapping is shown in Fig. S10A. 1, 2, 3, 4 means that the $T_1$ measurements were performed successively in the same position in different environments, where 1, 3 corresponds to air conditions and 2, 4 to water conditions. Each time the diamond is in a different environment, it is necessary to wait for at least 12 hours for the molecular conformation to change sufficiently.

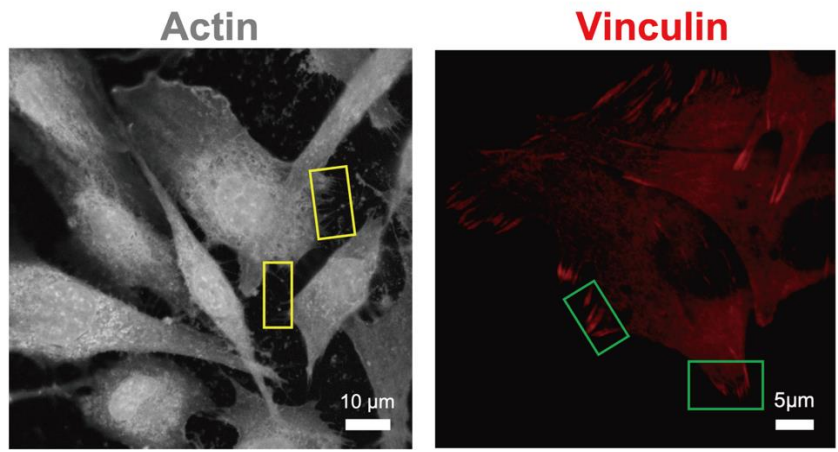

**Fig. S11.**

Representative fluorescence images of F-actin and vinculin of NIH 3T3 staining after $T_1$ measurements. Yellow boxes indicate pseudopodia and green boxes indicate focal adhesions, respectively. According to the images, the length of ~3 μm can be defined as the cell edge region (Fig. 5, A to C marked i, ii, iii, iv).

**Cell culture**

NIH 3T3 fibroblasts (ATCC) were cultured in standard DMEM (Gibco, 11965092) supplemented with 10% bovine growth serum (Gibco, 16030074) and 1% penicillin/streptomycin (Gibco, 15140122) at 37°C with 5% $CO_2$.

According to the stability test of the functional coatings, NIH 3T3 cells were seeded on the functionalized diamond slides (3 mm × 3 mm × 0.25 mm, Element Six, Optical Grade) for 16 hours to 7 days, followed by optical or fluorescent images acquisition.

**Immunofluorescence staining and microscopy**

Cells were washed once with cell culture medium and twice with PBS before fixation with 4% paraformaldehyde at room temperature for 15 min. Samples were then washed three times with PBS. Cells were permeabilized with 0.25% v/v Triton-X 100 in PBS for 10 min at room temperature, then washed three times with PBS. Nonspecific antibody adsorption was blocked by incubating samples with 1% w/v bovine serum albumin in PBST (0.1% v/v Triton-X 100 in PBS (PBST)) at room temperature for 45 min. Following primary antibody incubation (1:100, vinculin, Thermo), samples were washed twice with PBST and three times with PBS. Samples were then incubated with secondary antibodies, phalloidin 488 (Abcam, 1:1000) and DAPI at room temperature for 1 h, followed by washing three times with PBS. Immunofluorescence images were acquired and analyzed via confocal microscope (Zeiss710).

**Statistic assay**

Data of the cell adhesion study, measurement of $T_1$ and coating thickness are represented as mean ± standard deviation (S.D). Group differences were conducted by one-way ANOVA. P-values < 0.05 were considered statistically significant (*$p < 0.05$, **$p < 0.01$, ***$p < 0.001$, ****$p < 0.0001$). All statistical analyses were performed with GraphPad Prism 8.

**Simulation**

## The physical model of calculating Gd's influence on the $T_1$ of NV Center

From Fermi's Golden rule, we can estimate the relation rate of NV center (*79*):

$$\frac{1}{T_1} = \frac{1}{T_1^{bulk}} + 3\gamma_e^2 B_\perp^2 \frac{\tau_c}{1+\omega_0^2\tau_c^2}$$

where $T_1^{bulk}$ is the $T_1$ of NV in the bulk diamond. (We set it as 900 ns in this simulation based on experiment data.), $\omega_0$ is the NV zero-field splitting, where $\frac{\omega_0}{2\pi} = 2.87 GHz$ (*80*).

$\gamma_e$ is the gyromagnetic ratio obtained from the experiment.

$$\tau_c = \frac{1}{R_{Gd,tot}}, \text{ where } R_{Gd,tot} = R_{dip,Gd} + R_{vib} + R_{trans} + R_{rot}.$$

$$\hbar R_{dip} = \sqrt{\sum_{i\neq j}\langle H_{ij}^2\rangle} = \frac{\mu_0 \gamma_e^2 \sqrt{6} C_s}{4\pi}\left(\sum_{i\neq j}\frac{1}{r_{ij}^6}\right)^{\frac{1}{2}}$$

$R_{trans}$ and $R_{rot}$ are due to the Brownian motion. $R_{trans}$ is negligible.

$$R_{rot} = \frac{k_B T}{8\pi a^3 \eta f_r}, \text{ where } f_r = \left(\frac{6a_s}{a} + \frac{1+\frac{3a_s}{a+2a_s}}{\left(1+\frac{2a_s}{a}\right)^3}\right)^{-1}$$

$a_s$ and $a$ are the molecule radius of the solution and the Gd molecule. Here, we set $a_s = 0.14\ nm,\ a = 0.39\ nm$.

$$B_{\perp,i}^2 = \langle B_{x,i}^2\rangle + \langle B_{y,i}^2\rangle = Tr\{\rho(B_{x,i}^2 + B_{y,i}^2)\} = \left(\frac{\mu_0 \gamma_e \hbar}{4\pi}\right)^2 C_s \frac{2+3sin^2\alpha_i}{r_i^6}$$

$r_i$ and $\alpha_i$ are shown in Fig. S12.

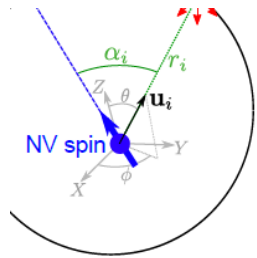

**Fig. S12.**

The schematic diagram of the model.

**The determination of simulation parameters**

**a. The density and depth of NV centers:** The same bulk diamond as used in other work has been adopted in our experiments (*56*). The density and depth of NV centers within the bulk diamond are 1000/um$^2$ and 5 nm, respectively.

**b. The density and depth of Gd$^{3+}$ molecules:**

The initial z-location of the Gd$^{3+}$ is determined by the Flory model of the PEG. For the PEG we use (Mw: ~1000 g/mol), the Flory radius of the PEG is $R_F = N^{\frac{3}{5}} \cdot l \approx 2.25$nm, if we also take the radius of Gd$^{3+}$ molecule (about 0.51 nm) into consideration, the Gd$^{3+}$ molecule is 2.76 nm away from the diamond surface at their free state. Based on the T$_1$ value we measured after complete force-responsive polymer modified diamond, the Gd$^{3+}$ molecule density is set to 9000/μm$^2$ in order to match the experimental data.

**The effective interaction range between $Gd^{3+}$ molecules and NV centers**

For a single NV center, we need to consider the influence of adjacent $Gd^{3+}$ molecules. In the calculation, we take the $Gd^{3+}$ molecules inside a circle region around NV centers into consideration. We call the region effective interaction range. Based on the calculation (Fig. S13), for a single NV center, if the interaction range is larger than the effective interaction range, the additional $Gd^{3+}$ will not influence the longitudinal relaxation process of the NV centers. Based on this, for each NV centers, we only consider influence of the $Gd^{3+}$ molecules in the effective interaction range of the NV centers.

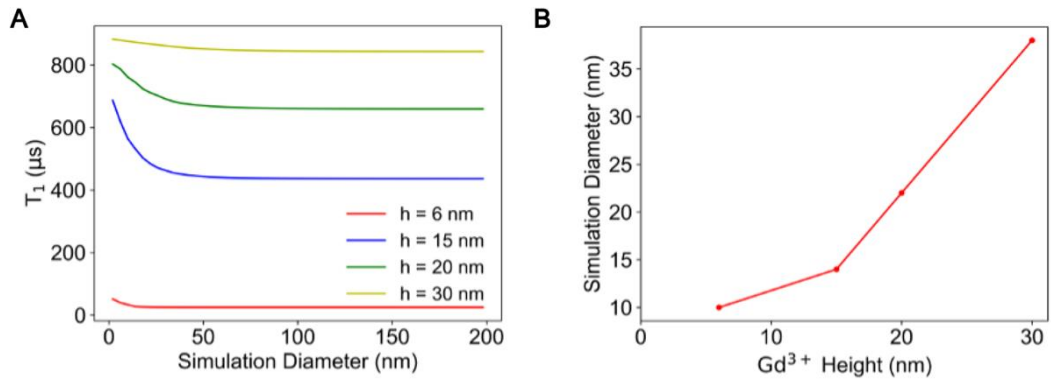

**Fig. S13.**

Determination of simulation range. **(A)** $T_1$ change as the simulation diameter increases. **(B)** Effective interaction range changes with the height of $Gd^{3+}$.

**The PEG model:**

**a. Theory model and experiment data for PEG**

For the extension process of the PEG, the Worm-Like Chain model (*55*) is suitable to describe it. The previous study shows that the result of the experiment performed in PBS buffer, which is the same buffer we use in our experiment, can be well described by the model (*81*).

**b. The PEG in our experiment**

In our experiment, the molecular weight of the PEG we use is 1000, and the Persistence Length of PEG is about 3.7 Å (*81*). For PEG (Mw: ~1000 g/mol), the number of subunits is 22.3, the net length of PEG is 0.278 nm ~ 0.358 nm, then the Counter length of the PEG (Mw: ~1000 g/mol) is 6.20 nm ~ 7.98 nm. Here, we choose 7.68 nm. Based on this, the relationship between force and extension can be theoretically described (Fig. S14).

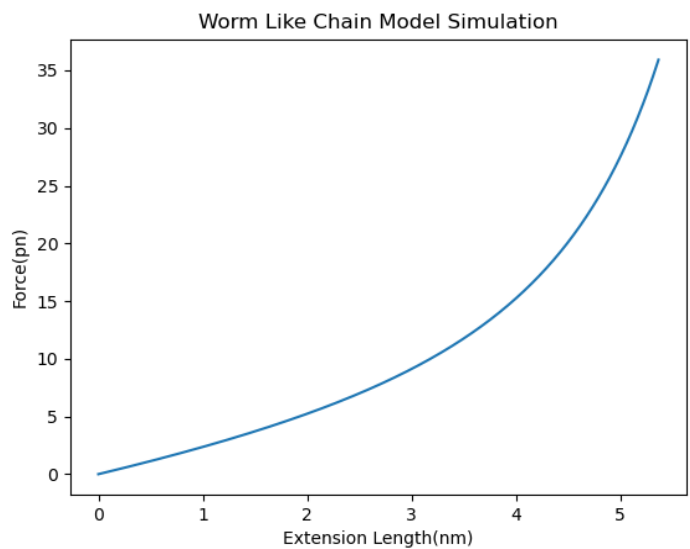

**Fig. S14.**

Force-Extension relationship of PEG (Mw: ~1000 g/mol) based on Worm-Like Chain model.

### c. Flory model of the PEG

When PEG is in a good solution, it can be described by the Flory model (*82*). The free length of it is Flory radius, which is:

$$R_F = N^{\frac{3}{5}} \cdot l$$

For the PEG we use in this experiment, the Flory radius is 2.25 nm, which means that when there is no external force applied on PEG, the extension of it is 2.25 nm.